\documentclass{webofc}
\usepackage[varg]{txfonts}   
\begin{document}
\title{Thoughts about the utility of perturbative QCD in the cores of neutron stars -- contribution to a roundtable discussion on neutron stars and QCD}
%
%

\author{\firstname{Aleksi} \lastname{Kurkela}\inst{1}\thanks{\email{aleksi.kurkela@uis.no}} 
}

\institute{Faculty of Science and Technology, University of Stavanger, 4036 Stavanger, Norway}

\abstract{%
In this contribution, I discuss the utility that perturbative QCD offers in studying the matter in the cores of neutron stars. I discuss the reasons \emph{why} perturbative QCD can constrain the equation of state at densities far below the densities where we can perform controlled calculations. I discuss \emph{how} perturbative QCD can inform nuclear modelling of neutron stars and how it influences equation-of-state inference. And finally, I discuss the implications to the QCD phase diagram and argue that interesting features in the equation of state revealed  by the QCD input may be used to argue for the existence of quark-matter cores in most massive neutron stars. 
}
\maketitle
\section{Introduction}
\label{intro}

The extremely rapid evolution in neutron-star measurements has turned the compact stars into a laboratory to study the properties of the densest matter in the observable universe. 
The properties of matter in the cores of the stars reflect the QCD phase diagram in the high-density region that is currently not under theoretical control \cite{deForcrand:2009zkb}. 

At sufficiently high (baryon number) densities, nuclear matter melts to quark matter \cite{Shuryak:1980tp}. 
This is an inevitable consequence of the asymptotic freedom of QCD. 
It is not currently known, however, how high the densities need be for quark matter to be formed. 
Whether or not these densities are realised in the cores of stable neutron stars is unknown; whether the conditions in post-merger remnants of a binary-neutron-star mergers are extreme enough is unknown. 
It is a grand question of nuclear astrophysics to determine whether or not deconfined quark matter is formed in compact stars. 

The equation of state of neutron-star matter is quantity of great interest  because it reflects the phase diagram QCD (e.g. \cite{Annala:2019puf,Fujimoto:2022ohj,Kojo:2021ugu,Tan:2021nat}). 
In many systems across various energy scales, different phases of matter have qualitatively different equations of states. 
Most dramatically, strong first-order transitions are sore discontinuities in the equation of state, and even a cross-over with a shift of active degrees of freedom can leave its qualitative mark in the equation of state. 
Therefore, if we were to be able to accurately determine the equation of state, we might at least indirectly be informed about the phase of matter in the cores of stars. 

The equation of state of neutron-star matter is a quantity of great interest also because it can be measured --- at least indirectly.  
It determines many of the macroscopic properties of the stars, such as the famous relation between masses and radii, as well as many other features, such as tidal deformabilities, mass shedding limits, oscillation frequencies and others. 
In this way, neutron stars are microscopes --- or rather \emph{femtoscopes} --- that map the QCD physics of the microscopic fermi-meter scale to the macroscopic scales of the stars.
And because of this mapping between the micro- and macrophysics, the neutron-star observations can be used to determine the equation of state.  
And therefore the observations of neutron stars may tell us about the phases of the matter in the cores of the stars. 

Lastly, the equation of state of neutron-star matter is quantity of great interest because studying it can enable us test our assumptions. 
When inferring the equation of state from astrophysical observations, many assumptions need to be made. 
Some of them, such as the number of parameters describing a shape of a hot spot on a X-ray star, are for the sake of convenience but some of them are about the fundamental physics we use to describe the compact stars. 
The fundamental assumptions are that neutron stars are described by Standard Model of Particle Physics and General Relativity. 
If the assumptions are right --- and if we can deal with all the technical complications --- a single equation of state should describe all the neutron stars, irrespective of whether they are quiescent or accreting, or undergoing a violent coalescence. 
Inconsistencies in the inferred equations of state from different systems and observables would flag an inconsistency in the assumptions, and can, in the most dramatic case, be a sign of physics beyond Standard Model and General Relativity. 
Similar logic of course applies to almost any other physical system, but none of them are as extreme as neutron stars, motivating us to at least to attempt to use them as a tool of discovery for new fundamental physics \cite{Bernitt:2022aoa}. 

All these above reasons strongly motivate to gain as much information about the equation of state as possible. And not only as accurately as possible but also redundantly, so that different systematic uncertainties and model assumptions and extrapolations involved in the inference using various observations can be validated. 

\begin{figure*}
\centering
\includegraphics[width=0.49\textwidth]{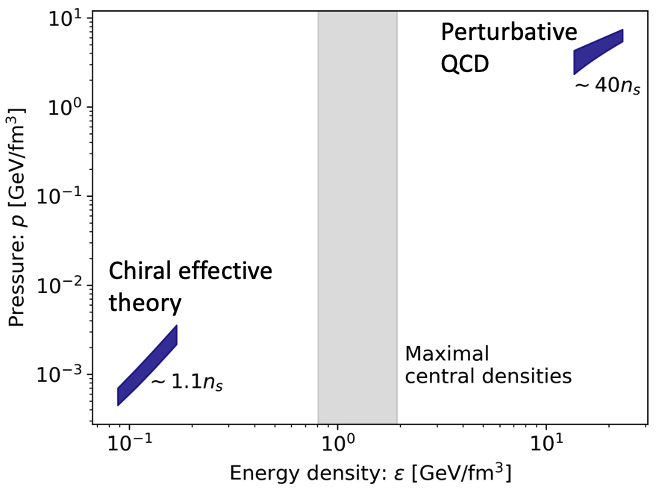}
\caption{The equation of state can be probed using chiral effective field theory at low densities, up to and around the saturation density $n\sim n_s$. At high densities $n\sim 40 n_s$, perturbation theory becomes quantitatively reliable. Cores of neutron stars have densities between these two limits.}
\label{fig-1}       
\end{figure*}

\section{Perturbative QCD}

In addition to observational inputs, the equation of state can also be constrained theoretically. 
Lacking the ability to theoretically access the equation of state in the interesting density regime of several saturation densities reached in the cores, $n \gtrsim 3 -8n_s$, we are limited to probe the equation of state at limits of low and high densities, see fig.\ref{fig-1}.
At low densities, the equation of state can be connected to nuclear scattering data and properties of nuclei using chiral effective theory by the virtue of the (approximate) chiral symmetry of QCD \cite{Bernitt:2022aoa}.
At high densities, the equation of state  can be connected to high-energy collisions using QCD perturbation theory by the virtue of asymptotic freedom \cite{Freedman:1976ub}. 

In both of these strategies, the theory calculations transport information from one physical system to another, from scattering to the equation of state. The inputs in chiral effective theory are the various low-energy coefficients of the effective theory.  In the case of perturbative QCD, they are the QCD coupling constant and quark masses, accurately measured in various experiments. 

It may be self-evident that the low-energy nuclear-physics calculations probing the equation of state up-to and around the nuclear saturation density provide interesting information about the neutron stars. 
What may be less evident, is to what extent the perturbative-QCD calculations that become quantitatively reliable only at the extreme densities of around $n\sim 40 n_s$ \cite{Freedman:1976ub, Kurkela:2009gj,Kurkela:2014vha} can inform us about neutron stars reaching only densities that are an order of magnitude smaller. 
It is clear that in some "moral" way it is interesting to know where the equation of state ends up at high densities. 
But does the knowledge of this limit actually tell something hard and non-negotiable at neutron star densities?

Among the first works on model independent equation-of-state inference were \cite{Hebeler:2013nza} and \cite{Kurkela:2014vha} (see also \cite{Read:2008iy}).
Both of these works attempted to infer the equation state in the interesting intermediate densities using large ensembles of piecewise polytropic functions (i.e. piecewise straight lines in $e$-$p$ --plane) that were constrained by demanding that each equation of state needed to be able to support a 2-solar-mass neutron star, given that such a star had been recently observed \cite{Demorest:2010bx}. The crucial difference between these two works was that \cite{Hebeler:2013nza} used only the low-density constraints, extrapolating the equation of state from saturation density up to core densities. Instead, in \cite{Kurkela:2014vha} the equation of state was interpolated between $1.1 n_s \lesssim n \lesssim 40n_s $ connecting the low- and high- density limits.
Already in these early works, a difference between extrapolation and interpolation could be seen (though some details of the implementations also differ making an apples-to-apples comparison not completely possible).  
At high densities, the interpolated equations of state soften compared to the extrapolated ones, limiting highest pressures attainable for a fixed energy density.

This softening is seen again and again in subsequent works. It is seen in works by myself and my collaborators as well as other groups that have imposed the perturbative-QCD input to their inference setups \cite{Annala:2017llu, Annala:2019puf, Annala:2021gom, Altiparmak:2022bke,Ecker:2022xxj,Jiang:2022tps,Marczenko:2022jhl}. The obvious criticism to the findings of the early works was that the interpolation over two orders of magnitude was performed with only two- or three-segment interpolation functions, it was performed only with single basis function and the ensembles where only rather small, and so on. Over the years these concerns have been addressed by myself and others, substantially increasing the number of segments, varying the interpolation functions and studying ensembles of enormous sizes. New observations have been included, and Bayesian methods have been included to ensure a correct propagation of errors. By-and-large the works that include the pQCD constrain see a softening of the equation state around $750$MeV/fm$^3$ in the equation of state and different quantities derived from the equation of state. The works without QCD do not see the softening (e.g. \cite{Somasundaram:2021clp}). It is almost as if the softening was caused by the perturbative-QCD input, not by the details of the interpolation. But how could that be? Isn't the perturbative QCD  reliable only an order-of-magnitude higher densities than what is reached in neutron stars?

\begin{figure*}
\centering
\includegraphics[width=0.45\textwidth]{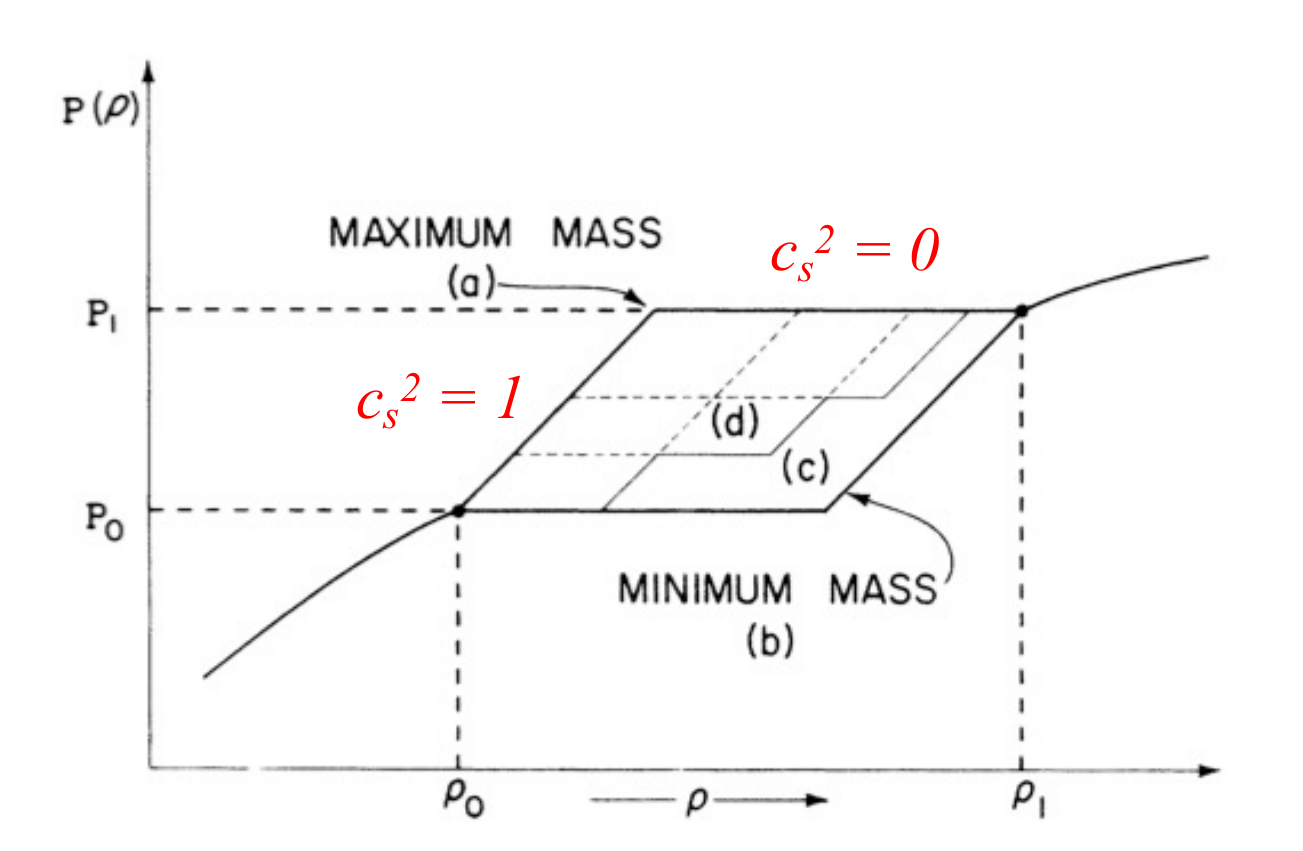}
\includegraphics[width=0.45\textwidth]{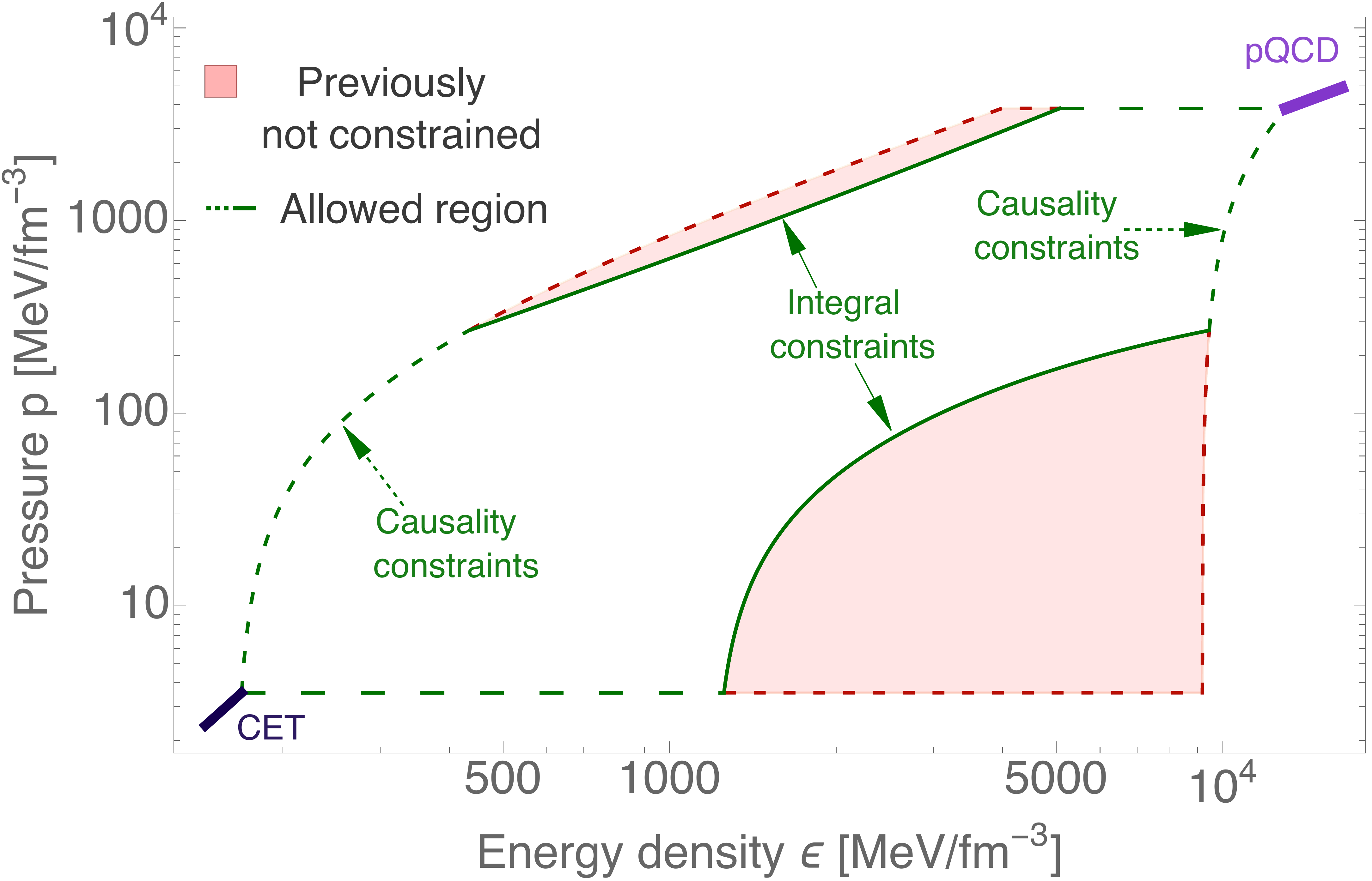}
\caption{(Left) The robust equation of state constraints arising from knowing the high-density limit of the reduced equation state by Rhoades and Ruffini, figure from \cite{Rhoades:1974fn}. Any equation of state connecting the low- and high-density limits with a causal and stable interpolation will necessarily lie inside the rhomboid. (Right) The robust equation of state constraints of \cite{Komoltsev:2021jzg}, requiring stability, causality and thermodynamic consistency. The dashed lines correspond to the rhomboid of Rhoades and Ruffini (in logarithmic scale), the "integral constraints" are due to requirement of thermodynamic consistency. }
\label{fig-2}       
\end{figure*}

\section{Robust equation-of-state constraints}
It has been known since the 70's that the information of the high-density equation of state does constrain the equation of state at lower densities \cite{Rhoades:1974fn}. 
The equation of state is not just any function, especially so in a relativistic context, and it has to fulfil certain general physical conditions. 
Conditions such as the requirement for mechanical stability and causality, i.e. having subluminal speed of sound given by $c_s^2 = \partial p/\partial \epsilon  < 1$. 
Rhoades and Ruffini discussed what are the most extreme equations of state that connect a known low-density limit to a high-density limit so that the resulting equation of state is still mechanically stable and causal at all densities. 
The most extreme equations are made of segments of maximally stiff $c_s^2 = 1$ equation of state and phase transitions. 
They form a rhomboid, inside which the true equation of state has to lie --- assuming that it is stable and causal at all densities.
Through the rhomboid,  the high-density limit constrains the equation of state at lower densities in a completely robust way, completely independent of the details of interpolation functions one might choose. 
Note that early works \cite{Hebeler:2013nza} and \cite{Kurkela:2014vha} also required stability and causality and therefore the latter one with the high-density input lies by construction inside the rhomboid. 

Based on this construction (and a more modern version \cite{Oter:2019rqp}), it did not seem like perturbative QCD could actually constrain the equation of state at lower densities without further assumptions. It seemed that it is always possible to connect anything to the high-density limit by means of large phase transition at very high densities. If so, then the ostensible constraints seen in the studies including perturbative QCD would be "soft" in the sense that by allowing more and more complicated and extreme interpolations the effect of QCD should finally go away. Using modern values of the high-density limit, QCD would give a strong constraint only above energy densities of $\sim 4$GeV/fm$^3$.

The construction from Rhoades and Ruffini did not, however, take into account all the information that we have at hand. 
In fact their most extreme equations of state are too extreme and are excluded along with a large class of other equations of states. 
Rhoades and Ruffini considered only the reduced form of the equation of state, energy density as a function of pressure $\epsilon(p)$. 
But this reduced form of equation of state has to arise from a consistent interpolation of an underlying grand canonical potential $\Omega(\mu)$ (or some other thermodynamic potential), from which the reduced equation of state can be derived using the normal thermodynamic relations. 
The thermodynamic potential --- to which we have the full access in the microscopic calculations in the two limits --- however contains more information than the reduced equation of state. 
It knows about the baryon densities, unlike the reduced equation of state. 
The equations of state that make the borders of rhomboid of Rhoades and Ruffini cannot be derived from a thermodynamic potential that consistently interpolates between the two limits. 

In  \cite{Komoltsev:2021jzg}, with Oleg Komoltsev, we generalised the construction of Rhoades and Ruffini to include the remaining condition and found the set of most extreme equations of states that are allowed by mechanical stability, causality and thermodynamic consistency. 

In this case, there are no single most extreme equations of state that would alone draw the borders of allowed region in the $\epsilon$-$p$ --plane. 
Instead, the extremal points arise from an infinite family of equations of state that each touch only at one point the boundary beyond which no equation of state is allowed to wonder.  
These additional constraints arising from the thermodynamical consistency are shown in \ref{fig-2} (right). 
They are denoted by "integral constraints" and the parts of the $\epsilon$-$p$ --plane that are newly excluded are marked with red color. 
What is striking about these constraints is that they reach down to very low densities, starting to constrain the equation of state as low as 2.3$n_s$. 

\begin{figure*}
\centering
\includegraphics[width=0.45\textwidth]{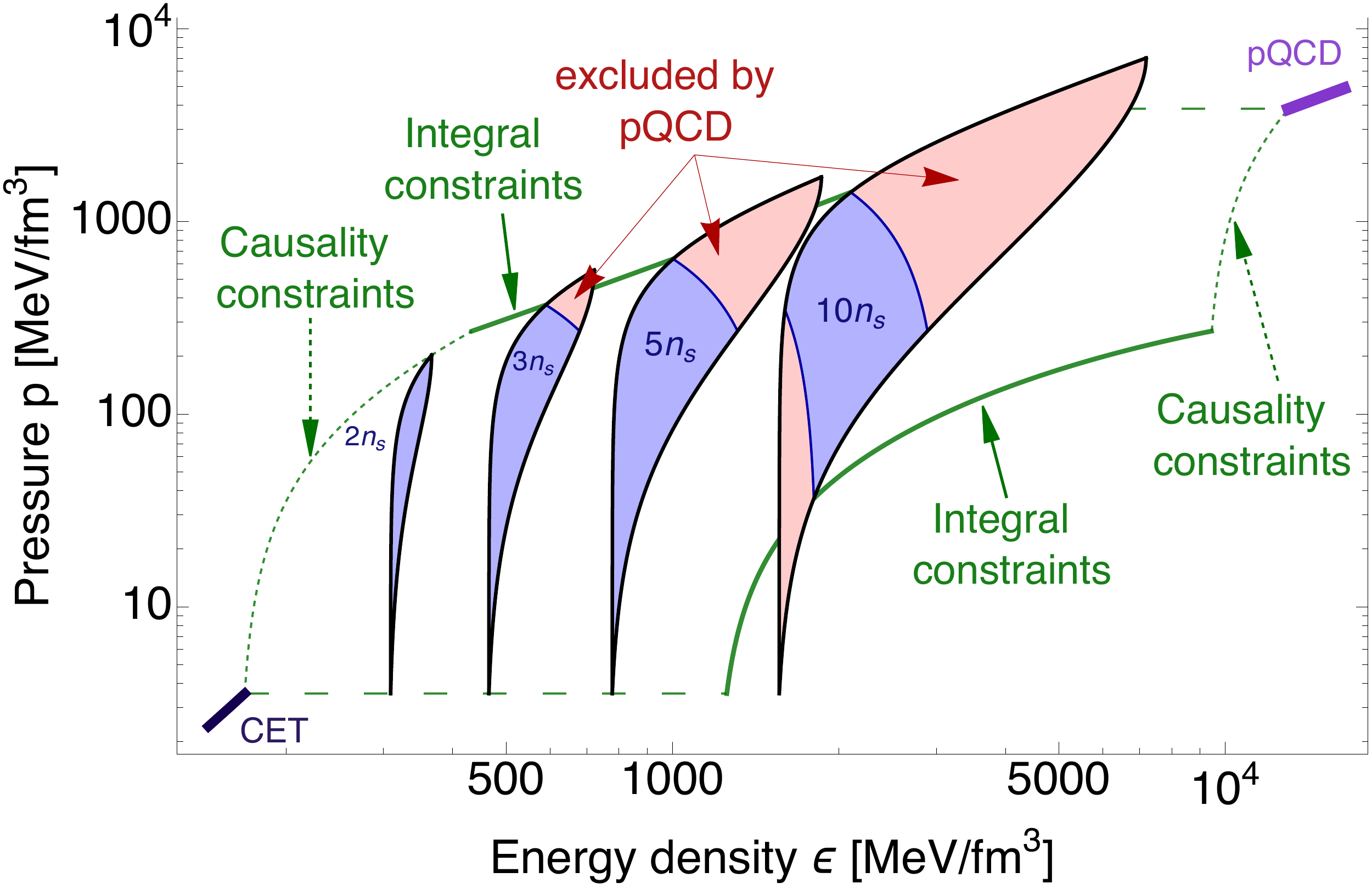}
\hspace{1cm}
\includegraphics[width=0.375\textwidth]{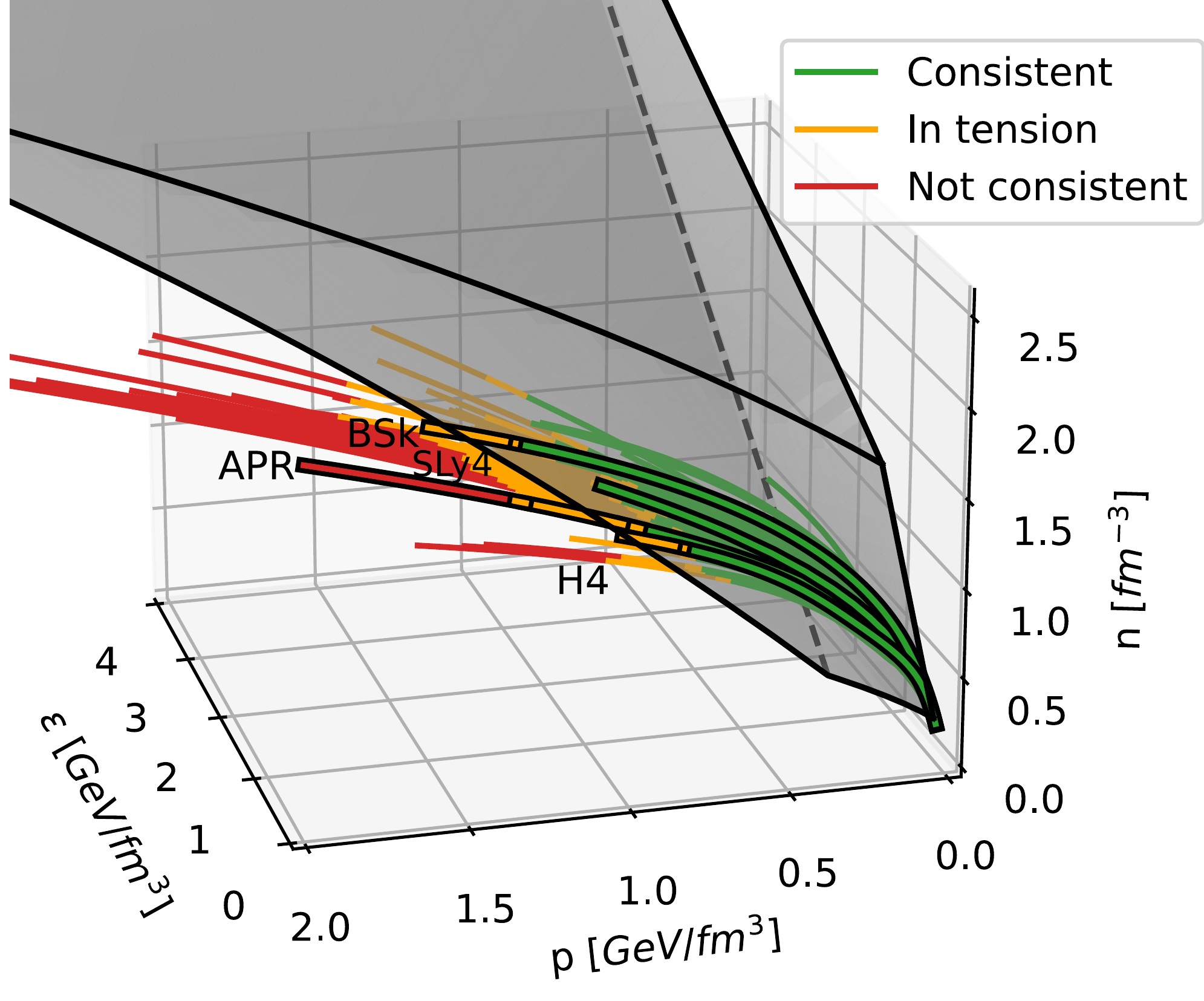}
\caption{Robust equation-of-state constraints from \cite{Komoltsev:2021jzg}. (Left) Regions of allowed values of energy density $\epsilon$ and pressure $p$ for fixed baryon number densities $n = 2,3,5$ and $10 n_s$. At five saturation densities, the
QCD input removes 75\% of the otherwise allowed values of $\epsilon$-$p$ (note the log-scale in figure). (Right) Comparison of the robust constraints with a large set of model equations of state from the ComPOSE database. Nearly all models violate the constraint at some density provided in the database. Equations of states outside the shaded volume are firmly excluded by QCD. Line colours correspond to the renormalization scale uncertainty of the perturbative calculation, for details see the original paper.}
\label{fig-5}       
\end{figure*}

The true utility of the constraint is seen when the particle number density is also included, as is shown in Fig.~\ref{fig-5} (left). The figure shows the allowed regions of $p$ and $\epsilon$ for fixed baryon densities, and demonstrates, e.g.,  how at $n =5 n_s$, large regions of values are excluded by the high-density input. In fact, in linear $\epsilon$-$p$ --plane the excluded area corresponds to 75\% of the otherwise allowed $\epsilon$-$p$ values. (Note that the figure \ref{fig-5} (left) is in log-log scale). 

 It is important to note that the constraints are not novel, in the sense many interpolation studies, starting from \cite{Kurkela:2014vha} and including many others \cite{Annala:2017llu, Annala:2019puf, Annala:2021gom, Altiparmak:2022bke,Ecker:2022xxj,Jiang:2022tps,Marczenko:2022jhl}, have included the QCD constraint by interpolating the thermodynamic potential, not the reduced equation of state. The novelty of this work is two-fold. Firstly, it clearly exposes \emph{why} the QCD calculation at very high density can offer information at low densities and demonstrated that this information is robust. And secondly, it offers the most conservative possible way to propagate this information from the high densities to neutron star densities without entangling the robust constraints with the inevitable details and choices related to the interpolation functions connecting the equation of state between two orders of magnitude in density. Any explicit interpolation will necessarily be more restrictive than the robust constraints. 
 
\section{Implications of robust QCD constraints to nuclear models}

It is of course possible that the constraints that arise from QCD exclude only equations of state that are so wild that they would anyways not be considered for one reason or another. 
A comparison between the constraint and a large number of equations of states that are typically used in neutron-star modelling and numerical simulations shows this not to be the case. 
Figure \ref{fig-5} (right) shows a large set of model equations of states acquired from the ComPOSE database \cite{Typel:2013rza}, where many model equations of states are stored and used as a data product in further modelling. 
In the figure, the grey volume corresponds to the values the equation of state can obtain so that it still can be connected to the QCD at high densities. 
It is immediately obvious that many, in fact most, of the equations of state leave the grey volume at some density and eventually become incompatible with QCD.

This is of course not to say that the these equations of states are to be completely discarded and without merit.  In the figure, the lines extend to the densities provided by the database. 
It is obvious that a nuclear model that knows nothing about quarks will not give a good description of the physics when extrapolated to high densities, while it may
still be perfectly fine at lower densities, and perhaps at all densities reached in the neutron stars. And indeed, some of them become incompatible with QCD only at densities that are higher than those reached in stable neutron stars.  But also, some of them become inconsistent already in the stable neutron stars. 

It is furthermore worthwhile to remember that while the equations of state outside the volume are firmly excluded, also those equations that approach the boundaries have to surely be too extreme to actually be realised in the nature.  An equation of state that touches the boundary can have only segments of $c_s^2 = 1$ and needs multiple very large first order transitions to fulfil the QCD constraints (for the details how they must behave, see \cite{Komoltsev:2021jzg}). While these equations of states cannot be excluded with the same level of rigour and certainty as the ones which directly violate the constraint, one could, e.g., imagine that these bounds should certainly affect nuclear model building by giving direction at high densities beyond the naive binary question whether a given model equation of state is allowed or not.

\section{Implications to equation-of-state inference} 
In addition giving direction to model building, the perturbative-QCD constraint affects the model-agnostic inference of equation of state. This is implicitly visible already in the interpolation studies that have incorporated the QCD constraint by interpolating the thermodynamic potential \cite{Kurkela:2014vha,Annala:2017llu, Annala:2019puf, Annala:2021gom, Altiparmak:2022bke,Ecker:2022xxj,Jiang:2022tps,Marczenko:2022jhl}. As discussed before, these studies consistently give differing results compared to the studies without QCD.  But in these studies, the questions about the generality of interpolation functions, number of segments and other details convolute the conclusion (even if several works have seen approximate convergence as a number of interpolation segments, nicely documented, e.g.,  in \cite{Altiparmak:2022bke}). A doubt always remains to what extent is the constraint arising from limitations of the interpolation and to what extent they are a robust prediction of QCD.  

\begin{figure*}
\centering
\includegraphics[width=0.47\textwidth]{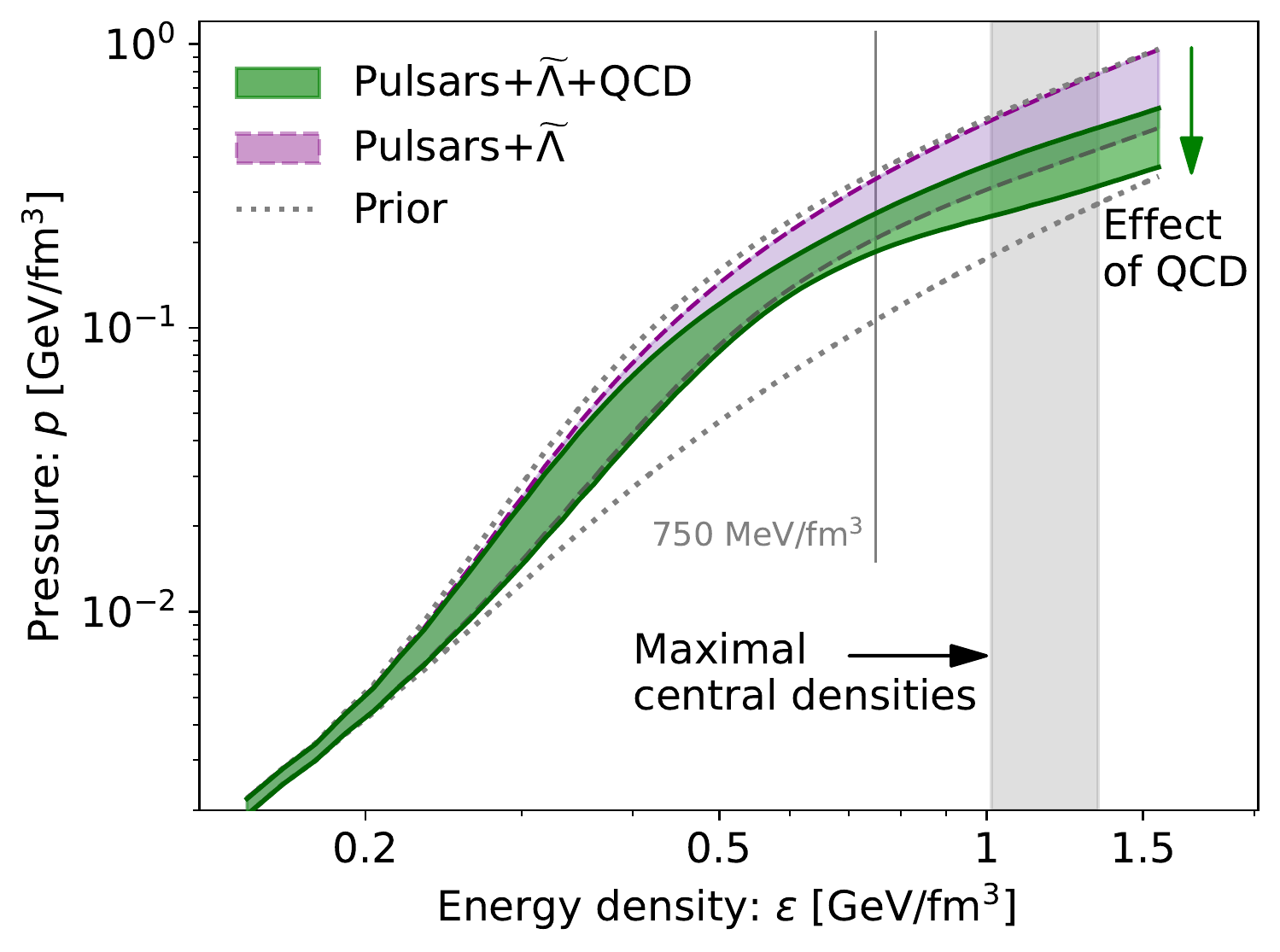}
\includegraphics[width=0.47\textwidth]{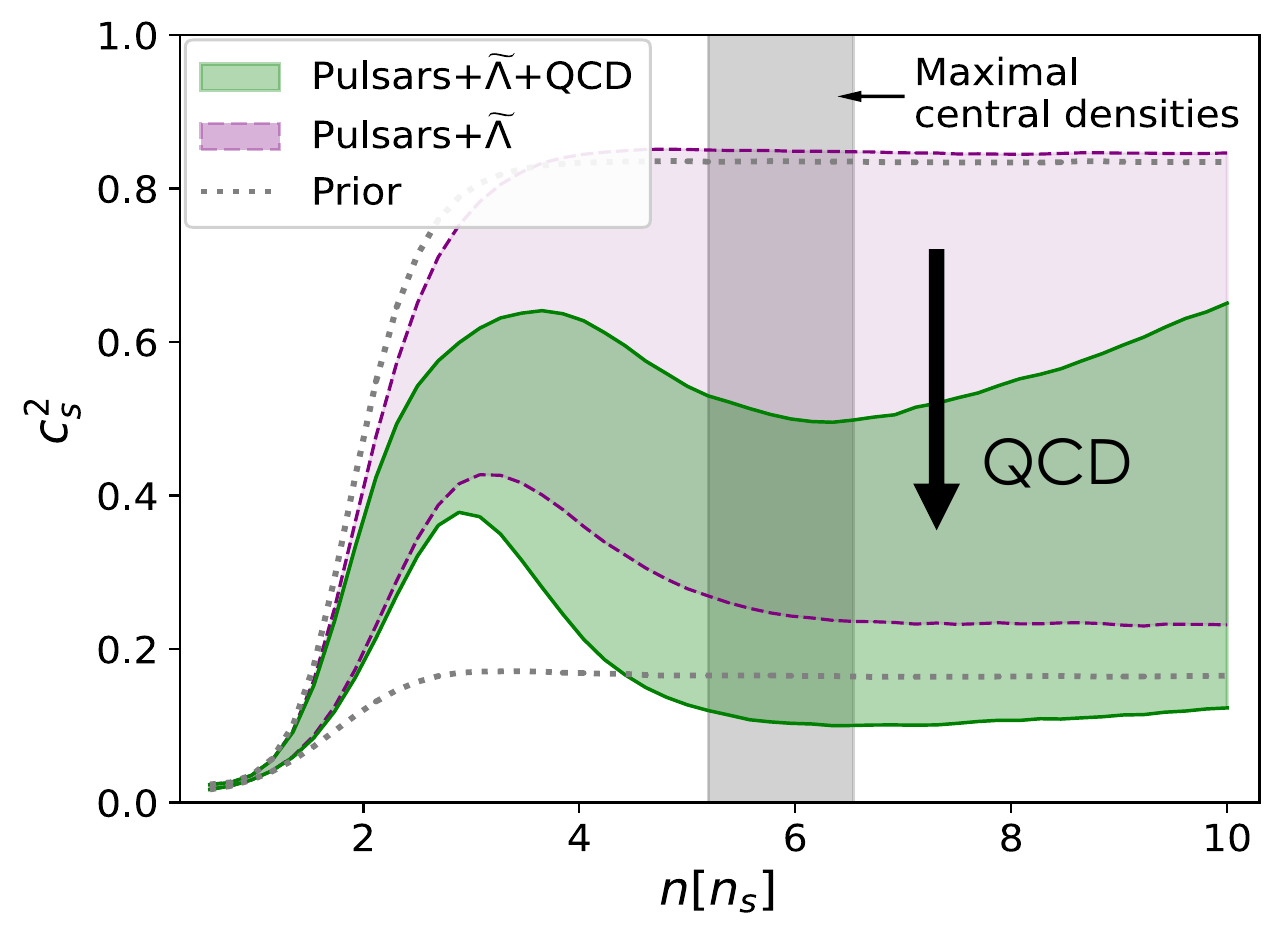}
\caption{The impact of the QCD input to the equation-of-state inference. (Left)The previously observed kink structure in the reduced equation of state around 750MeV/fm$^3$ appears when the QCD input in included. (Right) The kink feature represents it self as a peak in the speed of sound. It becomes visible only after the QCD input is included.   }
\label{fig-6}       
\end{figure*}

In \cite{Gorda:2022jvk}, our aim was to make the effect of QCD input to the equation of state inference explicit and disentangle it from the details of particular interpolation functions. In addition, the idea was to construct a setup where the QCD input could be turned on and off to cleanly see the effect that the robust QCD constraint has on the inference. 

Concretely, we constructed a setup that is similar to many other extrapolation setups that do not include the QCD constraint but with the option to impose the robust QCD constraints at the end.  Concretely, we did this by extrapolating the low-density equation of state to neutron star densities using and conditioning the prior ensemble with various neutron star measurements, ensuring that the equation of state is consistent with the current observations. Both the gaussian-process prior we used and all the measurements have been discussed elsewhere, and in this way the setup is one among many similar setups (see e.g. \cite{Essick:2019ldf, Landry:2020vaw}). The novel feature is, however, how we imposed the robust QCD constraint after the astrophysical observations. By doing so we could explicitly and robustly see the effect arising from QCD. We did this by constructing a likelihood function based on whether the endpoints of the extrapolation (at $n = 10 n_s$) could be connected to the QCD results with a stable, causal and consistent interpolation. In the construction of this likelihood function, we folded in the uncertainty arising from truncating the perturbative-QCD series at finite order using the standard practices from the high-energy community \cite{Duhr:2021mfd}.

The figure \ref{fig-6} demonstrates the effect of the QCD input to the equation of state inference. The results confirm indeed that the effect is to soften the equation of state at high densities, forming a knee feature at densities around $750$MeV/fm$^3$. The consequences of this interesting feature is also seen in the speed of sound, derived from the equation state, as a formation of a peak-like structure; the peak rises because of the need to be able to support a 2-solar-mass neutron star, and it has to come down because of the constraint posed by QCD \cite{Gorda:2022jvk} that constrains some (complicated) integral moment of the speed of sound.  Both of these features have been seen before in interpolation studies, but our work demonstrates that the result is indeed robust and does not arise from the choices involved in the interpolation. 

Note that the perturbative QCD value of the speed of sound $c_s^2 \approx 1/3$ does not enter at all in the derivation of the the robust QCD constraints. Still, the high-density part of the inferred equation of state is consistent with the perturbative value and it is easy
to imagine that the speed of sound could smoothly and continuously approach the perturbative value from below, as expected. 

One could wonder that if this feature indeed takes place at very high densities, can it be somehow observed? 
It does seem, for example, that the constraint coming from QCD has fairly small effect on the mass-radius relation. 
This is a double-edged sword. On one hand, it means that it would be very difficult to obtain the kind of information that QCD offers by mass and radius measurements making QCD a unique source of information, complementary to pulsar measurements. 
One the other hand, it makes the direct observational verification of the QCD predictions challenging. But it still may be possible.
One implication of the softening is to impose a limit on the maximal masses of neutron stars.  And through the limit in maximal masses, it implies that the final merger products in neutron star mergers (which are more massive objects than the progenitors) have to be black holes (not necessarily prompt, possibly via an intermediate hypermassive or supramassive stage), at least for certain mass configurations of the binary merger components. In particular, QCD implies that mass configuration of GW170817 should lead to a black hole as its merger product. On the contrary, without QCD, the model agnostic inference setups have no problem coming up with equations of state that would support a stable neutron star as the binary merger product of GW170817.

It is interesting that the electromagnetic counterpart of GW170817 is consistent with the formation of a black hole \cite{Gill:2019bvq, Rezzolla:2017aly,Margalit:2017dij} (perhaps through hypermassive neutron star phase), thus being consistent with the same conclusions that arises from QCD. But this conclusion arises from a very different reasoning, relying on different assumptions. In particular, the QCD argument does not rely on the complicated modelling of the central engine of the gamma ray burst, and the QCD argument does not require extremely sophisticated and delicate numerical modelling. 
That the conclusions agree leads to much desired redundancy of the argument. 

And on the flip side, an observation of stable binary merger product with a merger with certain chirp masses could falsify 
the QCD prediction and question the assumption that we can model neutron stars with Standard Model and General Relativity. 
Perhaps such events have already been recorded \cite{Jordana-Mitjans:2022gxy}? 

\section{What can the equation of state tell about the phase structure?}

\begin{figure*}
\centering
\includegraphics[width=0.47\textwidth]{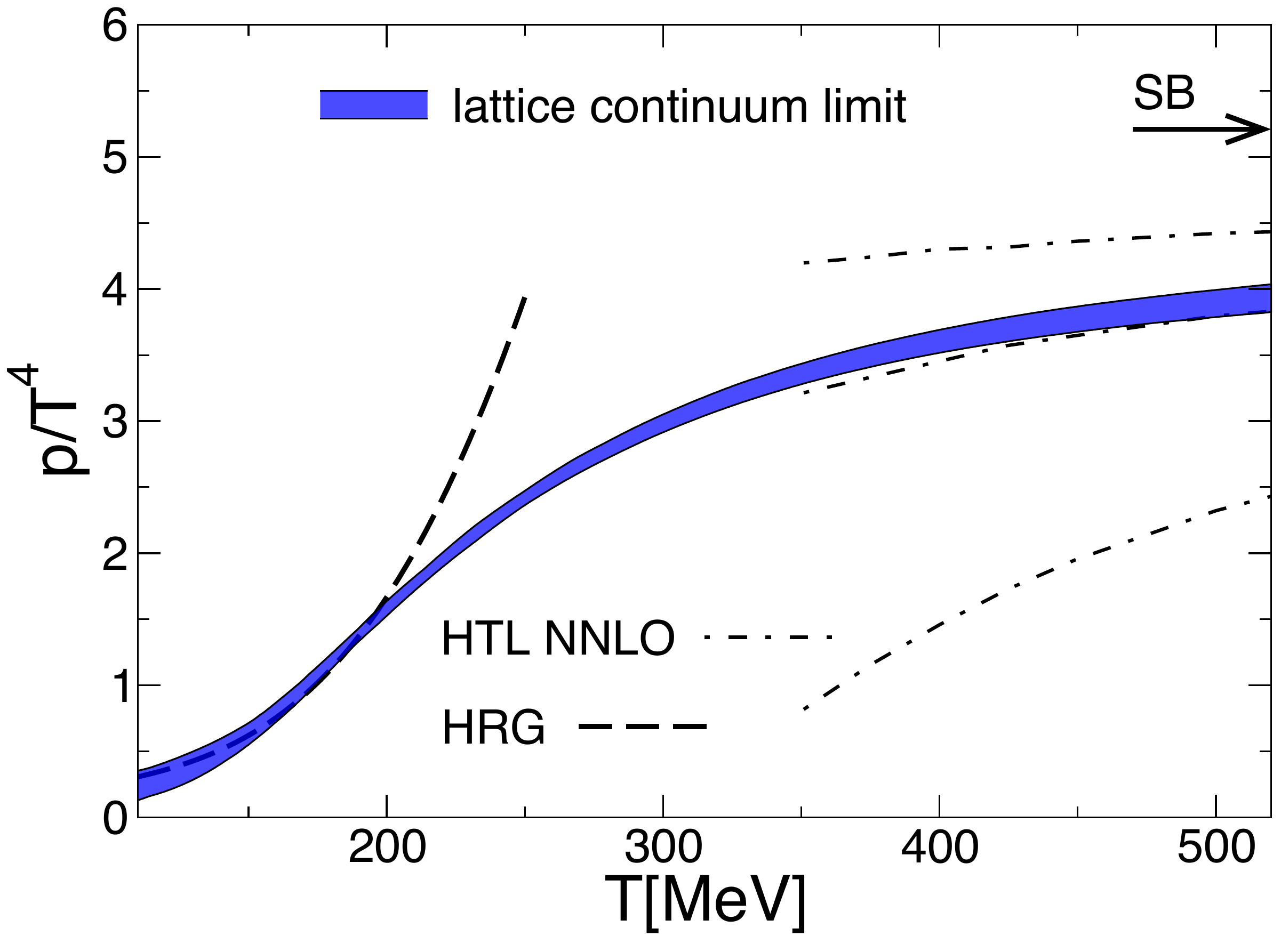}
\hfill
\includegraphics[width=0.47\textwidth]{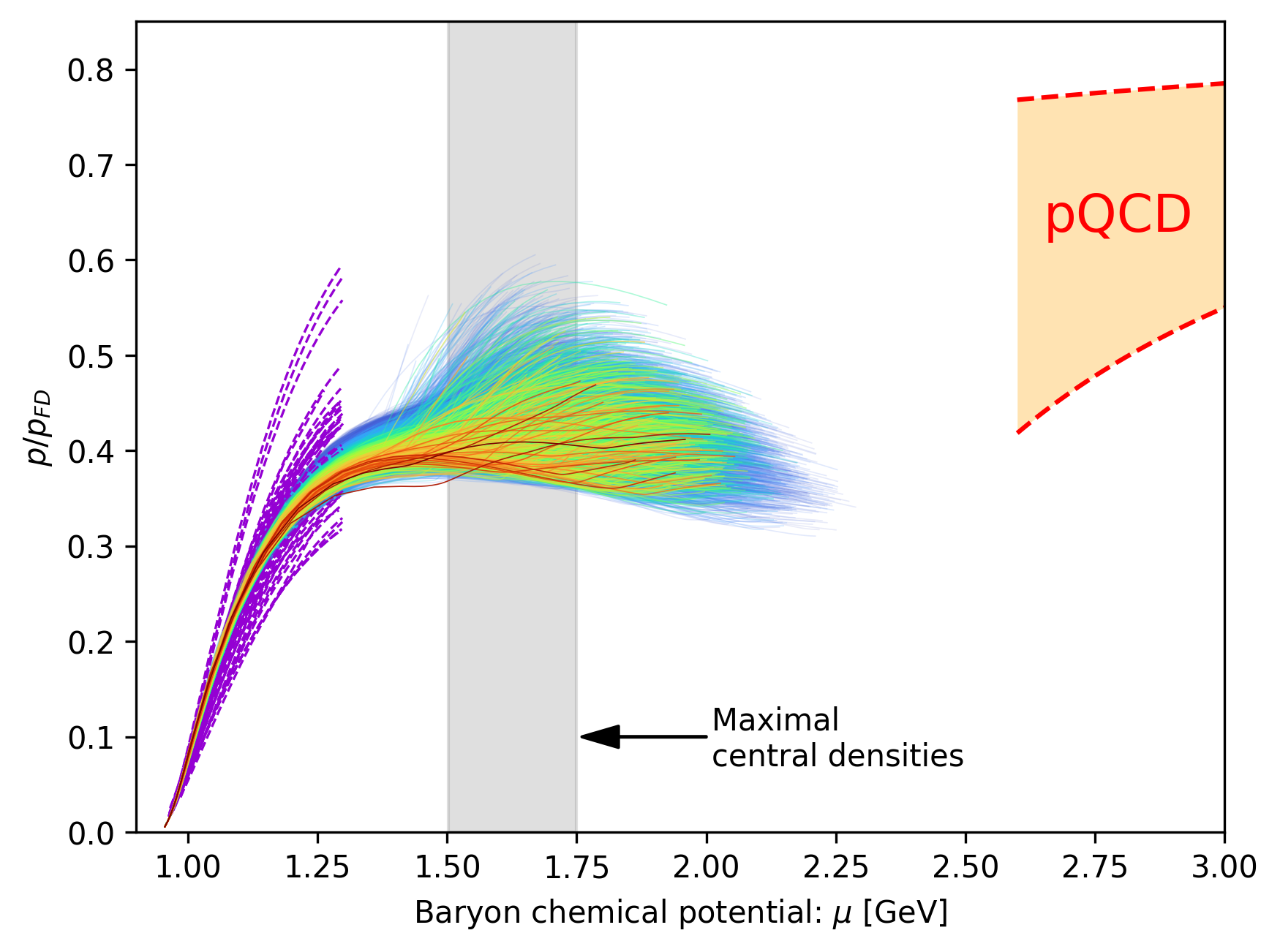}
\caption{(Left) The equation of state of hot quark-gluon plasma (at zero baryon-number chemical potential) determined by lattice field theory and tested in ultrarelativistic heavy-ion collisions.  The figure shows the pressure normalized by the free quark- and gluon gas. The lattice result is well reproduced by hadronic-resonance-gas model (HRG) at low temperatures and by perturbative (HTL NLO) calculation at sufficiently high temperatures. Figure from \cite{Borsanyi:2013bia}.
(Right) The analogous quantity at high baryon-number densities (zero temperature) ---  the pressure normalised by the the free-quark matter value. The colouring of the equations of state is based on their combined likelihoods; the lines terminate at fixed baryon-number density of $n=10n_s$. At low densities, the equation of state is reproduced by hadronic models (purple dashed lines). At high chemical potentials, the equation of state approaches the perturbative one. At the chemical potentials reached in cores of massive neutron stars, the equation of state seems to reproduce the qualitative features expected for quark matter. Figure adapted from \cite{Gorda:2022jvk}.
 }
\label{fig-7}       
\end{figure*}

It is an inevitable consequence of asymptotic freedom that at some high density, nuclear matter melts to deconfined quark matter. But does this happen inside neutron stars? And even if happens, how can it be diagnosed? There might or might not be a first order transition between the low-density hadronic phase and the quark matter phase. But it could also be that the two "phases" are separated by a smooth cross-over transition, as is the case at high temperatures, probed by the ultrarelativistic heavy-ion collisions at RHIC and the LHC. 

In heavy-ion collisions, the existence of quark matter (or quark-gluon plasma in this context) is well established. However, due to the cross-over nature of the transition, there is no single "smoking-gun" signal of quark matter. Rather, the evidence is built up from multiple observables and properties of the plasma, that at high enough temperatures support the picture of deconfined plasma (see e.g. \cite{Muller:2021ygo}). One of those quantities is the equation of state \cite{Gardim:2019xjs}. 

Non-perturbative computations in lattice field theory inform us about the equation of state at vanishing baryon number density \cite{Borsanyi:2013bia}. The equation of state has two separate domains. One at low temperatures, qualitatively characterised by the hadronic mass scales and at sufficiently low temperatures quantitatively reproduced by hadronic models (such as the hadronic resonance gas model \cite{Huovinen:2009yb}). And the other at high temperatures, qualitatively characterised by the (near-)conformality of the free quark- and gluon-gas, and at sufficiently high temperatures quantitatively reproduced by perturbative QCD. These domains are clearly, but continuously connected by the cross-over transition that takes place at pseudo-critical temperature around $T_c \approx 150$MeV (corresponding very roughly to energy densities of order $\sim 500$MeV/fm$^3$) . The value of the pseudo-critical temperature depends on the precise definition of it. But that this demarcation line is smooth and not clear-cut does not imply that the line does not exist. Below this line matter is hadronic, above it is quark matter. 

By empirically determining that heavy-ion collisions reach temperatures above the pseudo-critical one, we may argue that the matter created is quark-matter phase. It is not the only argument for deconfined matter in heavy-ion collisions. It may not be the strongest argument for deconfined matter in heavy-ion collisions. But it is one of the argument among others. 

In \cite{Annala:2019puf}, I suggested with my collaborators that a similar argument can be extended to quark matter in neutron stars. While in this case, we do not have the lattice to guide us, we may still examine the features in the empirically inferred equation of state.  With the "kink" in the inferred equation of state --- imposed by the interplay of the existence of massive neutron stars  and the QCD input --- the equation of state appears to have to separate domains. One at low densities, qualitatively characterised by the nucleon masses and at sufficiently low densities quantitatively reproduced by hadronic models. The other at high densities, characterised by the (near-)conformality of free quark matter, and at sufficiently high densities quantitatively reproduced by perturbative QCD. 

By determining that the cores of neutron stars reach this "conformalized" matter, reaching densities well above the kink, above the peak in the speed of sound, we may argue that the matter inside the cores of neutron stars is in quark-matter phase. 

Solutions to the Tollmann-Volkov-Oppenheimer equations seem to indicate that at least maximally massive stars likely reach densities in their cores where the equation of state conformalizes. It remains unclear if this happens in the cores of most massive stars observed so far, though. 

Similar conclusions have been reached also by other authors. For example, in a recent work \cite{Marczenko:2022jhl}, the authors study another variable, the conformal anomaly, derived from the equation of state, again making the point that the two phases have qualitatively different values of the conformal anomaly. By now, several works have come to the conclusion that the qualitative change in the equation of state exists, that it is indicative of transition to quark matter, and that it is likely reached within most massive neutron stars. 

A word of caution should be attached, however, as these conclusions do depend at least to some extent on the the details of the interpolation/extrapolation functions used in the analysis. These conclusions could change if the interpolation/extrapolation functions are allowed to have strong first-order transitions or other features. It may, for example, happen that a strong first-order transition destabilises the star and just when the quark matter would form, the star collapses in a catastrophic event \cite{Annala:2019puf}. Of course, the possibility that such an event could be observed is an exiting prospect (see e.g. \cite{Hempel:2015vlg,Sagert:2008ka}). 

Future neutron-star observations may make the features in the equation of state dramatically more pronounced. For example, there are neutron stars whose mass may be very large, such as PSR J0952-0607 whose mass was recently measured to be $M = 2.35 \pm 0.17 M_\odot$ \cite{Romani:2022jhd}. And there are very massive objects that may be neutron stars, such as the 2.6-solar-mass compact binary merger component of GW190814 of unknown nature \cite{LIGOScientific:2020zkf}. If either one of these turned out to be an extremely massive neutron star, it would have dramatic effect to equation of state inference --- especially so in combination with the QCD input. 
The observation of a very massive stable neutron star would lead into interplay between the required stiffness at lower densities and the QCD constraint that would force the equation of state have a very pronounced peak in the speed of sound, followed by a very soft equation of state up to high densities. This is exemplified by figure \ref{fig-7} which shows the impact that a discovery of 2.52 solar-mass neutron star would have on the inferred speed of sound; a less massive star would have a qualitatively similar impact, but I chose to show the very massive star here for a dramatic effect. 

\begin{figure*}
\centering
\includegraphics[width=0.47\textwidth]{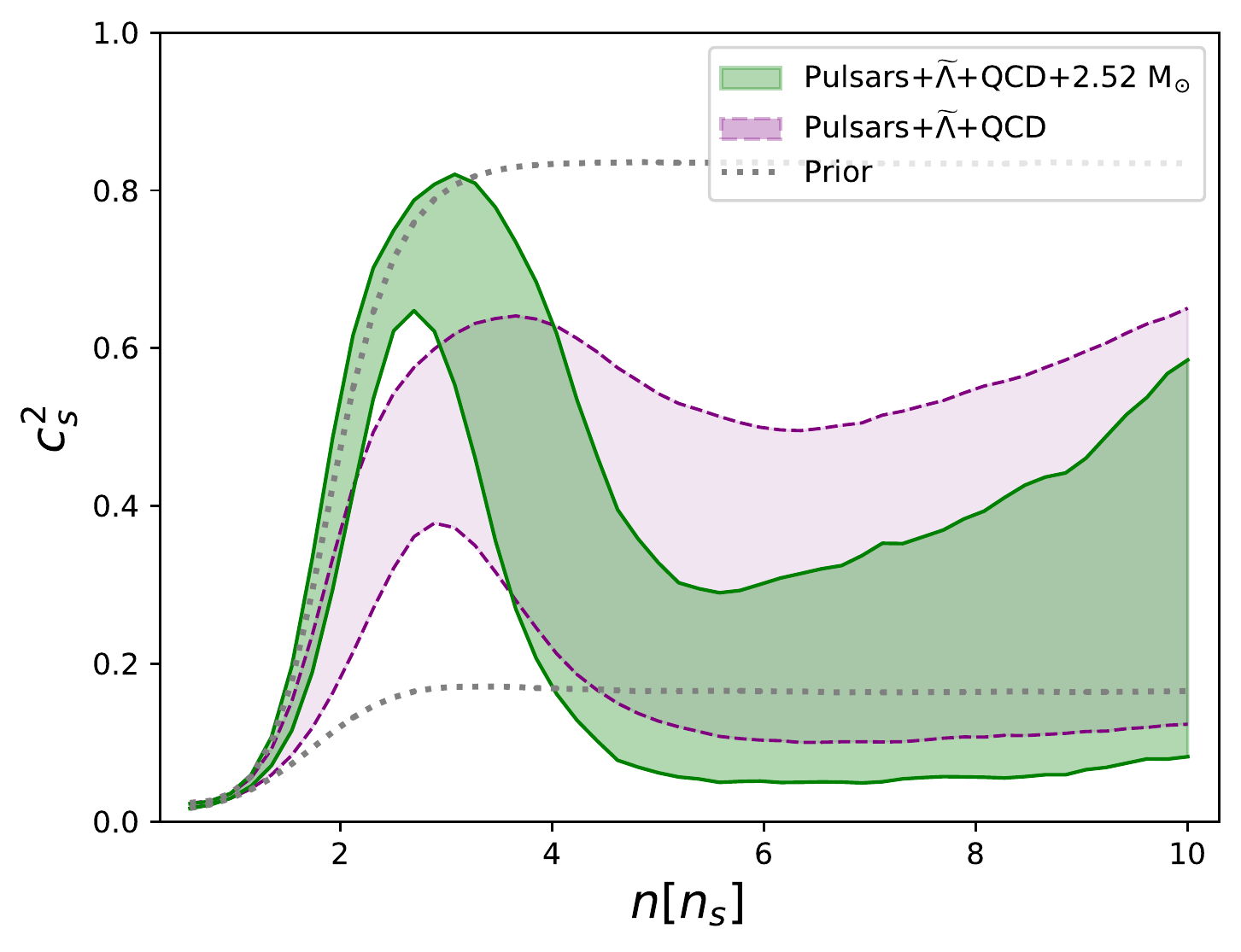}
\caption{The impact that a discovery of 2.52-solar-mass star would have on the inferred speed of sound. 2.52 solar masses corresponds roughly to the 1$\sigma$ upper limit of PSR J0952-0607 and lower limit of the binary merger component of GW190814. The purple band in the background corresponds to the results of \cite{Gorda:2022jvk}; the green band has additional requirement of stable 2.52-solar-mass neutron stars. The interplay between the required stiffness at lower densities to support a very massive star combined with the robust QCD constraint would force the equation of state to have a sharp peak. }
\label{fig-7}       
\end{figure*}

In the case of heavy-ion collisions the evidence of quark matter is composed of many quantities. In the case of neutron stars, at the moment, we have to rely on quantities derived from the equation of state. It would, naturally, be desirable to find other observables for which similar arguments could be made, both theoretically and observationally. Several suggestions have indeed been discussed, relating to the transport properties \cite{Alford:1997zt, Alford:2010gw, Alford:2007xm} as well as based on "smoking gun" signatures of phase transitions (e.g. \cite{Most:2018eaw, Fujimoto:2022xhv,Casalderrey-Solana:2022rrn}). In the end, to compellingly establish the existence of quark matter in the cores of neutron star, it is clear that we need --- like in the case of heavy-ion collisions --- evidence built up from multiple observables that all support the picture of deconfined quark matter. The features in the equation of state may be one of them.

\section{Conclusions}

The past decade has witnessed great advances in neutron-star observations turning them into a leading laboratory of ultra-dense matter. 
Aided by theoretical calculations, a quantitative picture equation of state at some reasonable level of accuracy is forming. 
And to consolidate and bolster this picture, redundancy of different inputs --- based upon different modelling assumptions and suffering 
from different systematics --- are needed. The assumptions and systematics of the QCD calculation are very clear: the assumption is
that nature is described by the Standard Model, and the uncertainties are quantified by the standard techniques of treating missing-higher-order
errors and renormalization scale uncertainty in quantum field theory.  In this way, information from QCD both complements and corroborates the astrophysical observations
that require significantly more extensive modelling to connect the observations to the equation of state.

This  quantitive picture suggests that the most massive neutron stars may well be harbouring cores made of quark matter. It remains a great challenge to further strengthen the arguments and to find new observables that could corroborate --- or disprove --- the existence of quark matter cores. This will require more work on neutron-star observations, and it strongly motivates further theoretical calculations, both at low and high densities.

The perturbative results that are currently used are at the Next-to-Next-to-Leading order level \cite{Freedman:1976ub, Kurkela:2009gj,Kurkela:2016was, Gorda:2021gha}, with parts of the calculation improved to N3LO \cite{Gorda:2021znl}. 
There is great motivation and push to bring the calculation to full N3LO. When this result will eventually be available, the uncertainties in the perturbative result will tighten, leading to tighter robust constraints at neutron star densities, and perhaps lead to quantitatively stronger argument for the existence of quark matter in neutron stars. 

Eventually, we would like to test the assumption that neutron stars are described by Standard Model and general relativity. Any equation-of-state inferences from astrophysical observations will be based on these assumptions and if they are not pertinent, the inferred equation of state will be incorrect. In contrast, the QCD calculations do not assume anything about the composition of neutron stars. By establishing a contradiction between the observations and the theoretical predictions, perhaps neutron stars can be used as tools for new fundamental physics. 

\section*{Awknowledgements}
I want to thank my co-panelists Kenji Fukushima, Michael Coughlin, and Nicholas Chamel for an interesting discussion organised and chaired by David Blaschke, whom I also of course thank. 
In addition, I thank my long-time collaborators Eemeli Annala, Eduardo Fraga, Tyler Gorda, Oleg Komoltsev, Joonas N\"attil\"a, Risto Paatelainen, Vasileios Paschalidis, Paul Romatschke, Saga S\"appi,  Juergen Schaffner-Bielich,  Aleksi Vuorinen and others for many fruitful discussions along these themes over the years.

\label{sec-1}

\bibliography{references}

\end{document}